\documentclass[runningheads]{llncs}
\usepackage{graphicx}
\usepackage{xcolor}
\usepackage[font=small]{caption}
\usepackage{float}
\usepackage{subcaption}
\usepackage{mwe}
\usepackage{hyperref}

\usepackage[utf8]{inputenc} 
\usepackage[T1]{fontenc}    
\usepackage{hyperref}       
\usepackage{url}            
\usepackage{booktabs}       
\usepackage{amsfonts}       
\usepackage{nicefrac}       
\usepackage{microtype}      
\usepackage{xcolor}         
\usepackage{amssymb}
\usepackage{algorithm,algorithmic,amsmath}
\usepackage{graphicx}
\usepackage{xcolor}
\usepackage{amsmath,bm}
\usepackage{graphicx}
\usepackage{textcomp}
\usepackage{xcolor}
\usepackage{bbm}
\usepackage{url}
\usepackage{wrapfig}
\usepackage{multirow}

\newcommand{\CR}{\mathcal{R}}

\DeclareMathOperator*{\argmin}{\arg\!\min}
\begin{document}
\title{Contribution Title\thanks{Supported by organization x.}}
%
%
\author{Wanyong Feng \and
Aritra 	Ghosh \and
Stephen Sireci \and Andrew S. Lan}
\authorrunning{F. Author et al.}
%
\institute{University of Massachusetts Amherst \\
Contact Email: \email{wanyongfeng@umass.edu}}
\title{Balancing Test Accuracy and Security in Computerized Adaptive Testing}

\maketitle              
\begin{abstract}
Computerized adaptive testing (CAT) is a form of personalized testing that accurately measures students' knowledge levels while reducing test length. Bilevel optimization-based CAT (BOBCAT) is a recent framework that learns a data-driven question selection algorithm to effectively reduce test length and improve test accuracy. However, it suffers from high question exposure and test overlap rates, which potentially affects test security. This paper introduces a constrained version of BOBCAT to address these problems by changing its optimization setup and enabling us to trade off test accuracy for question exposure and test overlap rates. We show that C-BOBCAT is effective through extensive experiments on two real-world adult testing datasets.
\end{abstract}

\section{Introduction}
Compared to conventional testing/assessment mechanisms, computerized adaptive testing (CAT) is a form of personalized testing that adaptively selects the next question based on students' answers to previous questions to reduce the test length effectively \cite{cat1,vats2013test,vats2016optimal}. A CAT system is generally comprised of the following components: a knowledge level estimator that estimates a student's current knowledge level given the answers to the previous questions, a response model that predicts how likely a student answers the question correctly given the knowledge level estimate and question features, and a question selection algorithm that selects the next most informative question given the output of response model \cite{cat3}. Despite the fact that CAT has been widely applied to real-world tests, an important limitation is that most of the existing question selection algorithms are static, which means that they cannot be improved over time as more and more students go through tests. Recently, researchers proposed several conceptual ideas for learning a data-driven question selection algorithm \cite{cat4}. 

\par
Bilevel Optimization-Based Computerized Adaptive Testing (BOBCAT) \cite{bobcat} (and extensions \cite{fake}) learns a data-driven question selection algorithm while supporting different response models. BOBCAT learns the question selection algorithm by solving the following bilevel optimization problem: the inner level optimization updates the student's knowledge level estimate using questions selected by the question selection algorithm, while the outer level optimization updates the question selection algorithm and response model by evaluating the latest knowledge level estimate on held-out data, i.e., student responses to questions not seen by the inner optimization. BOBCAT has several advantages over conventional CAT: i) BOBCAT can achieve the same knowledge level estimate accuracy with fewer questions selected, and ii) the question selection algorithm is agnostic of the response model, which enables the test administrators to explore the best combination that meets their needs.\par
Although BOBCAT has several advantages over conventional CAT, the learned question selection algorithm suffers from high question exposure and test overlap rates. The high question exposure rate means some questions are constantly being selected and the high test overlap rate means different students may receive similar test contents. This limitation could potentially affect the validity and fairness of the test by nefarious test preparation activities that focus on harvesting items administered to previous test takers \cite{issue}. Existing methods that address this issue are designed for traditional CAT settings \cite{ref2}, by either introducing randomness in question selection \cite{overlap} or using maximum clique algorithms \cite{ref1} and cannot be applied to data-driven CAT settings. 

\vspace{.2cm}
\noindent \textbf{Contributions.}
In this paper, we propose a constrained version of BOBCAT (C-BOBCAT) that can trade off test accuracy for question exposure and test overlap rates. C-BOBCAT i) uses a stochastic question selection algorithm instead of a deterministic one and ii) adds a penalty term to BOBCAT's optimization objective to promote the learned question selection algorithm to select diverse questions across different students. Therefore, we can trade off test accuracy for question exposure and test overlap rates via a temperature hyper-parameter. 
We validate the effectiveness of C-BOBCAT through experiments on two real-world adult test datasets and our implementation will be publicly available at: \url{https://github.com/umass-ml4ed/C-BOBCAT}.



\section{Methodology}
We first review BOBCAT and then detail the C-BOBCAT framework. 
\subsection{BOBCAT Background}
The goal of BOBCAT is to solve the following bilevel optimization problem:
\begin{equation}
  \underset{\bm{\gamma},\bm{\phi}}{\text{minimize}} \;\, \frac{1}{N}\!\textstyle\sum_{i=1}^{N}\!\sum_{j\in \Gamma_i}\!\ell \Big(\!Y_{i,j}, g(j;\!{\bm{\theta}}_i^{\ast})\!\Big) \!\!
  \label{eq:1}
\end{equation}
\begin{equation}
  \mbox{s.t.} \;\, {\bm{\theta}}_i^{\ast}\!=\!\!\argmin_{\bm{\theta}_i} \! \textstyle\sum_{t=1}^T \! \ell \Big(\!Y_{i,j_i^{(t)}}, g(j_i^{(t)};{\bm{\theta}_i})\!\Big)\!\!+\!\!\CR(\bm{\gamma}, {\bm{\theta}_i})\!\!
  \label{eq:2}
\end{equation}
\begin{equation}
  \mbox{where} \;\, j_i^{(t)} \sim \Pi (Y_{i,j_i^{(1)}}, \ldots, Y_{i,j_i^{(t-1)}}; \bm{\phi}),
  \label{eq:3}
\end{equation}
where $\Pi(\cdot)$ represents the question selection algorithm with parameters $\bm{\phi}$. The inputs are the responses of a particular student $i$ to the previously selected questions ($Y_{i,j_i^{(1)}}, \ldots, Y_{i,j_i^{(t-1)}}$), and the output is the next selected question index $j_i^{(t)}$. $g(\cdot)$ represents the global response model with parameters $\bm{\gamma}$. $\bm{\gamma}$ contains both the prior mean of student knowledge level estimate and question difficulties. The inputs are the student's knowledge level estimate ${\bm{\theta}}_i$ and question id $j$, and the output is the student’s likelihood of responding to the question correctly. The global response model can either be IRT-based or neural-network-based. Within a batch of $N$ students, BOBCAT splits each student $i$'s response questions into training part $\Omega_i$ and held-out meta part $\Gamma_i$.  For the inner-level optimization problem in (\ref{eq:2}), the question selection algorithm selects the next informative question from $\Omega_i$. The current student knowledge level estimate ${\bm{\theta}}_i^{\ast}$, which is the local adaption to the prior mean of student knowledge level estimate, is calculated as a function of $\bm{\phi}$ and $\bm{\gamma}$ by minimizing the summation of binary cross-entropy loss $\ell(.)$ on the selected questions and penalty function $\CR(\bm{\gamma}, {\bm{\theta}_i})$ that prevents large deviation. For the outer level optimization problem in (\ref{eq:1}), global response model and question selection algorithm are updated by minimizing the binary cross entropy loss $\ell(.)$ on the held-out meta data with the current student knowledge level estimate ${\bm{\theta}}_i^{\ast}$ as input. For further details, refer to \cite{bobcat}.
\subsection{C-BOBCAT} 
We now detail the modifications to BOBCAT. We turn BOBCAT's underlying question selection algorithm from deterministic to stochastic, which injects some randomness into the questions selected for each student. Specifically, we transform the original categorical question selection distribution to the Gumbel-Softmax distribution \cite{gumbel1,gumbel2} with a fixed temperature hyperparameter. Moreover, since the entropy of a distribution increases as the distribution approaches the uniform distribution, maximizing the entropy of the categorical question selection distribution can further encourage the learned question selection algorithm to select a diverse set of questions for each student. Specifically, we add the negative summation of the entropy of categorical question selection distributions of all selected questions for each student in (\ref{eq:5}) to the outer level optimization function in (\ref{eq:1}) to create the new outer level function in (\ref{eq:4}). 
\begin{equation}
  \underset{\bm{\gamma},\bm{\phi}}{\text{minimize}} \;\, \frac{1}{N}\!\textstyle\sum_{i=1}^{N}\!\sum_{j\in \Gamma_i}\!\ell \Big(\!Y_{i,j}, g(j;\!{\bm{\theta}}_i^{\ast})\!\Big) - \lambda H_i(\bm{\phi}) \!\!
  \label{eq:4}
\end{equation}
\begin{equation}
  \mbox{where} \;\, H_i(\bm{\phi}) = \textstyle\sum_{t=1}^{T}\! H\Big( \Pi (Y_{i,j_i^{(1)}}, \ldots, Y_{i,j_i^{(t-1)}}; \bm{\phi})\Big) \label{eq:5}
\end{equation}
During the training process, the model needs to maximize the combination of both prediction accuracy on the held-out meta data and the uncertainty of the question selection algorithm, which is reflected in the entropy regularization term. This balance is controlled by the value of the hyperparameter $\lambda$: when $\lambda = 0$, the problem reduces to the original BOBCAT bi-level optimization problem; when $\lambda = \infty$, the solution to the problem is a question selection algorithm that selects each question with equal probability, i.e., the entropy is maximized when the question selection distribution is uniform distribution \cite{entropy}. In practice, test administrators can explore different values of $\lambda$ and select a suitable value, to achieve a desirable balance between maximizing the accuracy of the test and maintaining acceptable question exposure and test overlap rates, according to the requirement of each testing scenario.

\section{Experiments}
We now detail experiments we conducted on two real-world adult test datasets to validate C-BOBCAT’s effectiveness.
\subsection{Data, Experimental Setup, Baseline, and Evaluation metrics}
We adopt two new publicly available datasets collected from the Massachusetts Adult Proficiency Test (MAPT), the reading comprehension test (MAPT-Read), and the math test (MAPT-Math). In both datasets, there are a total of more than 90K students, 1.7K questions, and 4M question responses \cite{mapt}. We use these datasets since they are collected in real tests while datasets used in the original BOBCAT paper are collected during longer periods of student learning. As a result, the MAPT datasets are closer to the actual real-world situation addressed by CAT, i.e., a test taker's ability roughly remains unchanged during a short test. We choose the gradient-calculation-based method from BOBCAT to learn the question selection algorithm with C-BOBCAT framework (\textbf{C-BINN-Approx} and \textbf{C-BIIRT-Approx}). We split the datasets over students into train, validation, and test sets with the ratio of $6 : 2 : 2$ respectively. For each student $i$, we partition the questions they answer (and their responses) into the training part ($\Omega_i$, $80\%$) and the held-out meta part ($\Gamma_i$, $20\%$). We choose the IRT-based active learning method (\textbf{IRT-Active}), and the IRT-based random question selection method (\textbf{IRT-Random}) as the baseline models. We use the area under the receiver operating characteristics curve (\textbf{AUC})  as metrics to evaluate the predictive performance on the held-out meta data $\Gamma_i$. We use the Scaled Chi-square Statistics of the question exposure rate (\textbf{EXPOSE-CHI}) and the average value of test overlap rate between every two students (\textbf{OVERLAP-MU}) to measure question exposure and test overlap rates \cite{expose}. 
\begin{figure}[tp]
    \centering
    \begin{subfigure}[b]{0.475\textwidth}
        \centering
        \includegraphics[width=\textwidth]{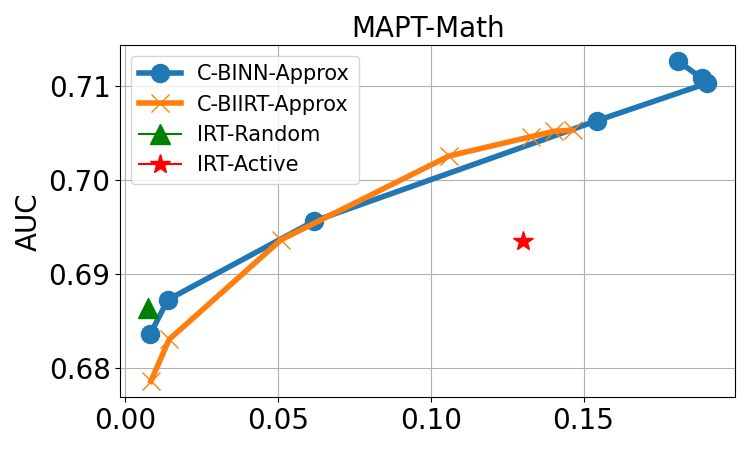}  
        \label{fig:mean and std of net14}
    \end{subfigure}
    \begin{subfigure}[b]{0.475\textwidth}  
        \centering 
        \includegraphics[width=\textwidth]{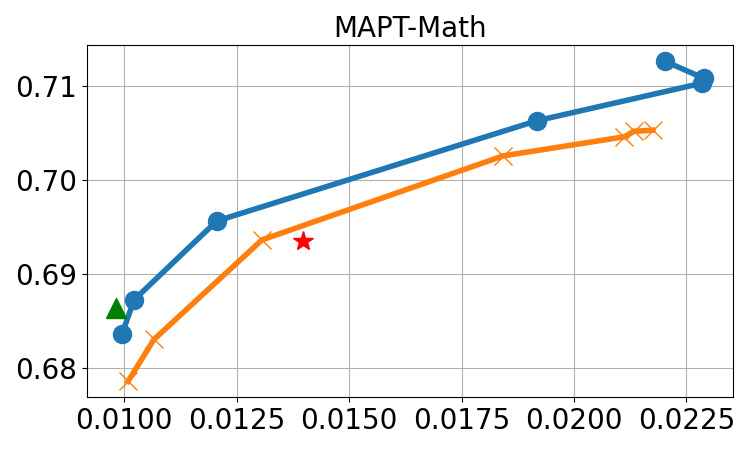}
        \label{fig:mean and std of net24}
    \end{subfigure}
    \vskip\baselineskip
    \begin{subfigure}[b]{0.475\textwidth}
        \centering 
        \includegraphics[width=\textwidth]{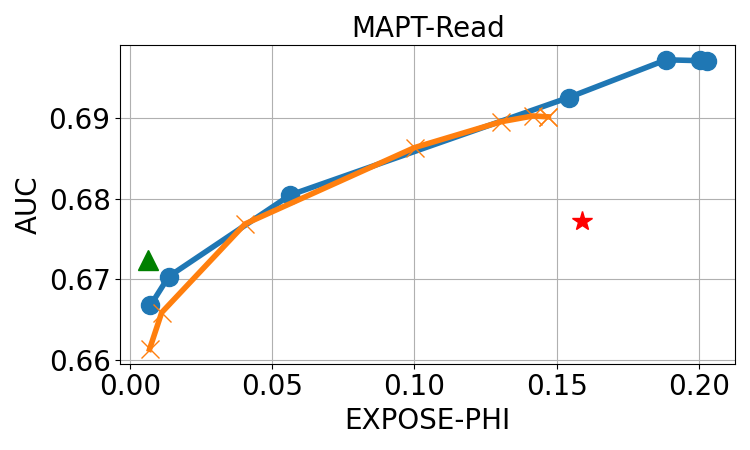}  
        \label{fig:mean and std of net34}
    \end{subfigure}
    \begin{subfigure}[b]{0.475\textwidth}
        \centering 
        \includegraphics[width=\textwidth]{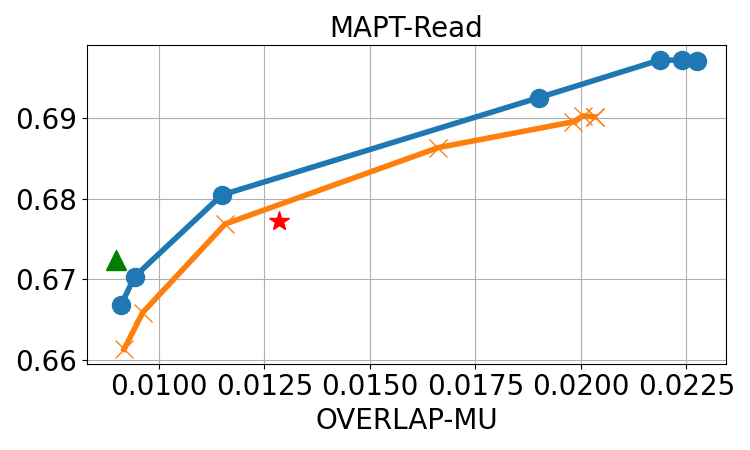}   
        \label{fig:mean and std of net44}
    \end{subfigure}
    \caption{C-BOBCAT can effectively trade off test accuracy (y-axis, AUC on held-out meta data) for test security (x-axes, EXPOSE-PHI and OVERLAP-MU) on both datasets.} 
    \label{fig:graph}
\end{figure}
\subsection{Results and Discussion}
In Figure~\ref{fig:graph}, we use a series of plots to demonstrate how C-BOBCAT trades off test accuracy (AUC on held-out meta data) versus test security (EXPOSE-PHI and OVERLAP-MU). For both C-BINN-Approx and C-BIIRT-Approx, each point on the curve corresponds to a specific value of the hyperparameter $\lambda$. The baselines correspond to single points since they do not support this tradeoff. We observe that AUC and EXPOSE-PHI are positively correlated in the first column of the graph. Both metrics reach the minimum as $\lambda$ approaches infinity, i.e., when the categorical question selection distribution's entropy is maximized. In this scenario, the question selection algorithm selects each question with equal probability, which decreases both the question exposure rate and test accuracy. Both metrics reach the maximum as $\lambda$ approaches zero, i.e., when the categorical question selection distribution's entropy is minimized. In this scenario, the question selection algorithm greedily selects the question that leads to the highest test accuracy, which also results in a high question exposure rate. A similar relationship can also be found between OVERLAP-MU and AUC in the second column of the graph. More importantly, with the same EXPOSE-PHI and OVERLAP-MU values, both C-BINN-Approx and C-BIIRT-Approx achieve higher AUC values than the IRT-Active baseline. This observation implies C-BOBCAT is agnostic of the underlying response model and always achieves a better balance between test accuracy and question exposure and test overlap rates than the IRT-Active baseline. 

\section{Conclusions and Future Work}
In this paper, we proposed C-BOBCAT, a framework that attempts to strike a balance between test accuracy and security in CAT settings, which we demonstrated to be effective via experiments on two real-world datasets. Avenues of future work include i) investigating the effect of combining questions' info with the responses from students as the new inputs to the question selection algorithm, ii) for multiple choice questions, investigating the effect of predicting the exact option that students select instead of correctness in both inner and outer level optimization, and iii) deploying C-BOBCAT in real-world tests. 

\section{Acknowledgements}
The authors thank the NSF (under grants 1917713, 2118706, 2202506, 2215193) for partially supporting this work. 

\bibliographystyle{splncs04}
\bibliography{ref}

\end{document}